\documentclass[aps,prl,preprint,groupedaddress,superscriptaddress,nofootinbib]{revtex4-1}
\usepackage{graphics}
\usepackage[mathcal]{euscript}
\usepackage[latin1]{inputenc}
\usepackage{latexsym}
\usepackage{amsmath}
\usepackage{hyperref}   
\usepackage{graphicx}   
\usepackage{verbatim}  
\usepackage{color}      
\usepackage{subfigure}  
\usepackage{hyperref}   
\usepackage{bm}
\usepackage{graphicx}
\usepackage{amssymb}
\usepackage{bbm}
\usepackage{float}
\usepackage{slashed}
\usepackage{amsfonts}
\usepackage{euscript}
\usepackage{mathrsfs} 
\usepackage{bbding}
\usepackage{pifont}
\usepackage{cancel}
\usepackage{amssymb,fge}
\usepackage{pifont}
\usepackage{multirow}

\usepackage[usenames,dvipsnames]{xcolor}
\hypersetup{colorlinks=true, citecolor=Cerulean, linkcolor=Cerulean,
urlcolor=Cerulean}

\usepackage{tikz}
\usetikzlibrary{decorations.pathmorphing,decorations.markings}

\begin{document}

\title{Observational evidence for the formation of trapping horizons\\ in astrophysical black holes}


\author{Ra\'ul Carballo-Rubio}
\email{raul.carballorubio@sissa.it}
\affiliation{SISSA, International School for Advanced Studies, Via Bonomea 265, 34136 Trieste, Italy}
\affiliation{INFN Sezione di Trieste, Via Valerio 2, 34127 Trieste, Italy} 
\author{Pawan Kumar}
\email{pk@astro.as.utexas.edu}
\affiliation{Department of Astronomy, University of Texas at Austin, Austin, TX 78712, USA}
\author{Wenbin Lu}
\email{wenbinlu@astro.as.utexas.edu}
\affiliation{Department of Astronomy, University of Texas at Austin, Austin, TX 78712, USA}


\bigskip
\begin{abstract}

Black holes in general relativity are characterized by their trapping horizon, a one-way membrane that can be crossed only inwards. The existence of trapping horizons in astrophysical black holes can be tested observationally using a reductio ad absurdum argument, replacing black holes by horizonless configurations with a physical surface and looking for inconsistencies with  electromagnetic and gravitational wave observations. In this approach, the radius of the horizonless object is always larger than but arbitrarily close to the position where the horizon of a black hole of the same mass would be located. Upper bounds on the radius of these alternatives have been provided using electromagnetic observations (in optical/IR band) of astronomical sources at the center of galaxies, but lower bounds were lacking, leaving unconstrained huge regions of parameter space. We show here that lower bounds on the radius of horizonless objects that do not develop trapping horizons can be placed using observations of accreting systems. This result is model-independent and relies only on the local notion of causality dictated by the spacetime geometry around the horizonless object. These observational bounds reduce considerably the previously allowed parameter space, boosting the prospects of establishing the existence of trapping horizons using electromagnetic observations. 

\end{abstract}

\maketitle

\section{Introduction}

The most distinctive elements of black holes (BHs) are the trapping horizon, located at the Schwarzschild radius in spherically symmetric situations, and the singularity at the center of the gravitational potential. In dynamical situations, the concept of trapping horizon captures the intuitive idea of the local boundary of the BH at a given moment of time \cite{Hayward1993,Gourgoulhon2008} (in technical terms, here we use the term trapping horizon to denote a three-dimensional submanifold foliated by marginally trapped surfaces, which is also known as dynamical horizon when space-like \cite{Ashtekar2004}). In stationary black holes, the trapping horizon is coincident with the event horizon \cite{Visser2014}.

There is no conclusive observational evidence for horizons in astrophysical BHs, but only partial indications \cite{Narayan1997,Narayan2002,McClintock2004,Narayan2008,Broderick2009,Lu2017,Cardoso2016,Cardoso2017} as summarized briefly in the following. A convenient way to illustrate this assertion is considering the theoretical exercise of replacing BHs with horizonless objects (HOs) with mass $M$ and a coordinate radius $r_0$ that is slightly greater than their Schwarzschild radius $R_{\rm S}=2GM/c^2$, namely $r_0=R_{\rm S} + \ell$ where $0<\ell\ll R_{\rm S}$ (note that we will be working in spherically symmetric situations for simplicity; our main results do not rely on this simplification, as it will be emphasized). For our purposes here we can remain agnostic about most of the properties of these objects (see \cite{Cardoso2017} for a recent review), as the ideas on which the central result of this paper is based only invoke certain model-independent considerations. A convenient measure of the coordinate radius of the HO is the dimensionless quantity \mbox{$\mu=(r_0-R_{\rm S})/r_0\simeq \ell/R_{\rm S}$}. As described below, compact enough HOs remain unconstrained (this remains true even if taking into account recent gravitational wave observations \cite{Cardoso2016}), which represents a clear reminder that our experimental knowledge of astrophysical BHs is still limited. Constraining the existence of HOs as defined here and showing the existence of trapping horizons in astrophysical BHs are two sides of the same coin.

\section{Upper bounds}

It is possible to derive upper bounds for the radius $\mu$ of HOs using electromagnetic observations \cite{Narayan1997,Narayan2002,McClintock2004,Narayan2008,Broderick2009,Lu2017}. Nearly all galaxies have a central massive object (CMO) of mass  about $10^6 - 10^{10}\ M_\odot$ \cite{Kormendy2013}, which are widely believed to be BHs \cite{Narayan2005}. Stars get tidally disrupted when passing within the Roche limit of the CMOs of mass $\lesssim 10^8\ M_\odot$ and produce bright optical/UV transients \cite{Hills1975,Rees1988}. Such tidal disruption events (TDEs) have been observed by various surveys carried out in the optical, UV and soft X-ray wavelengths, giving a TDE rate of $\sim 10^{-5}\ \mbox{yr}^{-1}\ \mbox{galaxy}^{-1}$ \cite{Gezari2007,vanVelzen2014,Holoien2015}. For CMOs of mass $\gtrsim10^8\ M_\odot$ the Roche limit approaches $R_{\rm S}$ and in that case an infalling star is swallowed as a whole, if the CMO has a trapping horizon. However, if the CMO has a physical surface (instead of a trapping horizon), then the infalling star collides with the surface, and the shocked baryonic gas forms a radiation-pressure supported envelope that shines near the Eddington luminosity for months to years (until $\sim 10^{54}\ \mbox{erg}$ -- the star's rest mass energy -- is radiated). The rate of such transients from star-CMO collisions over cosmic volume can be calculated by integrating the in-falling rate of stars over the mass function of CMOs. Such an emission has been ruled out by Pan-STARRS1 3$\pi$ survey \cite{Chambers2016} at 99.7\% confidence, if CMOs have a hard surface at radius larger than $\mu\simeq4\times10^{-5}$ \cite{Lu2017}. 

Due to its proximity, the CMO in our Galaxy, Sgr A*, gives a much stronger upper limit on $\mu$. Sgr A* is currently accreting at an extremely low level, with luminosity \mbox{$L_{\rm disk}\simeq10^{36}\ \mbox{erg}\ \mbox{s}^{-1}$} from the accretion disk, peaking at wavelength $\sim 0.1\ \mbox{mm}$ \cite{Narayan1997b}. This is about $10^{-9}$ times the Eddington luminosity for mass $M\simeq4\times10^6\ M_\odot$ \cite{Ghez2008}. The efficiency of the accretion disk at converting gravitational energy to radiation is less than 100\%, which suggests a lower bound on the accretion rate $\dot{M}\geq L_{\rm disk}/c^2 \sim 10^{15}\ \mbox{g}\ \mbox{s}^{-1}\sim 10^{-11}\ M_\odot\ \mbox{yr}^{-1}$. Instead of a trapping horizon, if Sgr A* has a hard surface at $\mu\ll1$ and the system is in steady state such that the escaping energy balances the infalling mass/energy, then the luminosity seen at infinity must be $L_\infty\simeq \dot{M}c^2 \geq 10^{36}\ \mbox{erg}\ \mbox{s}^{-1}$. The emission from the hard surface has a blackbody spectrum with temperature \mbox{$T_\infty=[L_\infty/(4\pi R_{\rm a}^2\sigma_{\rm SB})]^{1/4}\simeq 3.5\times10^3 (L_{\infty,36})^{1/4}\ \mbox{K}$}, where $L_{\infty,36}=L_\infty/(10^{36}\ \mbox{erg}\ \mbox{s}^{-1})$, $\sigma_{\rm SB}$ is the Stefan-Boltzmann constant, and we have used the apparent size $R_{\rm a} = 3\sqrt{3}R_{\rm S}/2$ for a spherically symmetric object (the dependence of $R_{\rm a}$ on spin is weak and hence neglected here). Thus the surface emission should be bright in the infrared (wavelength $\sim 1\ \mu\mbox{m}$). It has been shown that the observed infrared fluxes at $1-10\ \mu\mbox{m}$ from Sgr A* are one to two orders of magnitude below this prediction \cite{Broderick2009}, thus this surface emission does not exist. 

Still, one may argue that Sgr A* is not in a steady state of perfect balance between the rates of infalling and escaping energy \cite{Lu2017,Cardoso2017}. If the radiation from the accreted gas is produced at the radius $\mu=(1-R_{\rm S}/r_0)\ll 1$, then only a small fraction (roughly proportional to $\mu$) of the radiation escapes to infinity and the rest follows a highly curved trajectory that brings it back to the hard surface \cite{Abramowicz2002}. The time it takes for a photon to bounce $\mu^{-1}$ times (in order to escape) is $t_{\rm esc}\sim R_{\rm S}/\mu c$. A steady state can be reached only when the accretion timescale $t_{\rm acc}$ (that provides a measure of the time scale for the CMO mass to increase by about a factor of 2) is longer than $t_{\rm esc}$. For Sgr A*, a natural timescale for accretion is the Eddington timescale \mbox{$Mc^2/L_{\rm Edd}\simeq 4\times10^8\ \mbox{yr}$} ($L_{\rm Edd}$ is the Eddington luminosity), so we have \mbox{$t_{\rm acc}/t_{\rm esc}\sim4\times10^{14}\mu\,[t_{\rm acc}/(4\times10^8\ \mbox{yr})]$} (various evidences, such as the X-ray echoes from nearby molecular clouds, show that Sgr A* was likely accreting at a much higher rate in the past \cite{Koyama1996,Ponti2012}, which should give an even brighter surface emission than our conservative calculation). Therefore, we conclude that the lack of surface emission in the infrared constrains the location of a possible surface to $\mu\lesssim 3\times10^{-15}$ for Sgr A*.

\begin{figure}[h]%
\begin{center}
\includegraphics[width=0.6\textwidth]{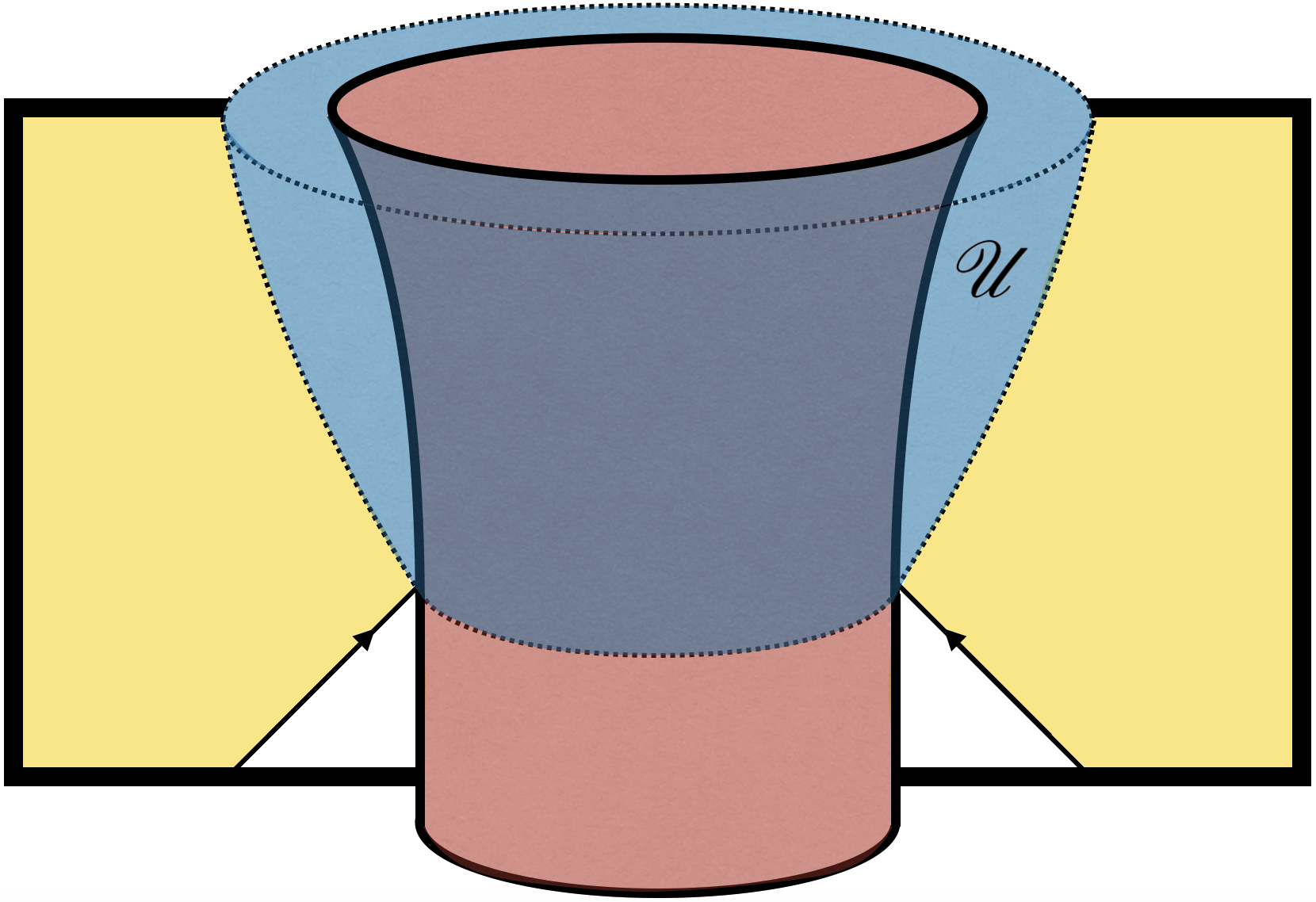}
\end{center}
\vspace{-0.5cm}
\caption{Schematic representation of the spacetimes in which our main theoretical result is formulated, with time flowing upwards. The white region corresponds to a vacuum BH geometry. The ingoing arrows separating the white and yellow regions indicate the onset of accretion. The internal cylinder-like red region marks the interior of the HO with all its model-dependent features, including a dense core and possibly an atmosphere of Planckian width. The blue cone in dashed lines delimits the maximal possible extension of the HO as determined by the causality condition, and it is defined geometrically in terms of outgoing null geodesics that depart from the boundary of the HO at the moment in which the first particles of accreting matter reach the latter. The region $\mathscr{U}$ may also contain accreting matter that has been backscattered. These regions are defined in general with no need of invoking any particular symmetries or specific properties of the accreting matter and the HO.}
\label{fig:fig1}%
\end{figure}%

It is clear that this upper bound alone can only serve as an indication, but not a proof, of the existence of trapping horizons \cite{Abramowicz2002}:  the proper radial length $\ell/\sqrt{\mu}$ can be as small as the Planck length $L_{\rm P}\simeq2\times10^{-33}\ \mbox{cm}$ and, in fact, this is its natural value in a number of scenarios \cite{Cardoso2017,Carballo-Rubio2018}. There are no known electromagnetic observations that can push down the upper bound on the size of the CMO in order to rule out well-motivated models in which $\mu \sim (L_{\rm P}/R_{\rm S})^\alpha$ with $\alpha\in[1, 2]$. In this parametrization, $\alpha=1$ for the choice $\ell = L_{\rm P}$, which gives $\mu\sim10^{-38}\times(M_\odot/M)$, while $\alpha=2$ corresponds to the choice $\ell/\sqrt{\mu}= L_{\rm P}$, giving $\mu\sim10^{-76}\times(M_\odot/M)^2$. Between the latter case ($\alpha=2$) and the upper bound derived above using data from Sgr A*, there is a huge parameter space for $\mu$ that spans more than 70 orders of magnitude. The possible values for $\mu$ are parametrized very compactly in terms of $\alpha\in(0,2]$ where $\alpha\rightarrow0$ would correspond to a HO with a radius that is roughly twice its Schwarzschild radius. Hence, it may seem impossible to constrain ultra-compact HOs using electromagnetic observations (and gravitational wave observations are still far from being able to probe the spacetime geometry so close to the horizon \cite{Cardoso2016,Cardoso2017}; in any case, these will provide again upper bounds only). 

\section{Causality considerations}

The CMOs we observe have matter and radiation falling onto them, and therefore their size is changing with time. If a physical surface exists, this infalling matter (that we are assuming to be standard matter, i.e., with positive energy density and pressure) has to be assimilated into the exotic matter that makes up the surface of the HO. Regardless of the physical details of this process, according to the fundamental paradigm of physical laws the relevant interactions must propagate at a finite speed (at most, with the local speed of light) that is determined by the properties of the accreting matter. As the growth of a given stellar structure follows from the interaction between the accreting matter and its surface, triggered when the first particles of accreting matter reach the latter, then the dynamical evolution of the surface must be along causal trajectories in spacetime. Every known astrophysical object satisfies this causality condition. It is important to stress that this condition is not in contradiction with superluminal apparent motions in special or general relativity, which are allowed with the caveat that no superluminal interactions or energy exchanges occur \cite{Landau1982}. This basic and model-independent requirement turns out to be surprisingly restrictive.

The geometric region in spacetime $\mathscr{U}$ that is defined as the maximal possible extension of the HO under accretion (Fig. \ref{fig:fig1}) is bounded due to the causality condition. Due to the stronger gravitational pull, it becomes smaller the more compact the HO was initially (e.g., the lower the value of $\mu$ in spherically symmetric situations). On the other hand, the flux of accreting matter into this region can take arbitrarily large values. Therefore, for any conceivable initial configuration it is always possible to form a trapping horizon by considering a large enough accretion rate $\dot{M}|_{\partial\mathscr{U}}$ \cite{Malec1991}. An immediate corollary is that more compact initial configurations would have lower values of the critical accretion rate leading to the formation of a trapping horizon. 

This argument is completely general, meaning that it is not based on assuming certain symmetries (such as spherical symmetry), or specific properties of the accreting matter or the HO. However, in order to extract relations between the different quantities and observables of interest, it is necessary to consider specific situations, which in most cases would require numerical analysis. Here we show that these relations can be obtained analytically in a simple (but still physically relevant, as explained in detail below) model, which assumes spherical symmetry and pressureless relativistic accreting matter. Then, the spacetime metric describing the geometry outside an accreting HO is given by the Vaidya metric \cite{Vaidya1951,Griffiths2009}:
\begin{equation}
ds^2=-\left[1-\frac{2GM(v)}{c^2r}\right]dv^2+2dvdr+r^2d\Omega^2.\label{eq:vaidya}
\end{equation}
The function $M(v)$ is the mass enclosed inside the radius $r_0(v)$ of the HO, where the geometry above has to be glued with the internal geometry describing the relevant properties of these hypothetical objects. Let us consider an idealized situation in which there is no accretion for $v < v_0$ or $v > v_{\rm f}$, but there is a constant accretion rate in the closed interval $v \in[v_0,v_{\rm f}]$, with $v_0 < v_{\rm f}$ real numbers. Then $M(v<v_0) = M$ and $M(v_0\leq v\leq v_{\rm f} )= M + (v-v_0)\dot{M}/c$, where $\dot{M}$ is the accretion rate measured by stationary observers at $r\rightarrow\infty$ in terms of their proper time \cite{Nielsen2010}. 

\begin{figure}[h]%
\begin{center}
\includegraphics[width=0.45\textwidth]{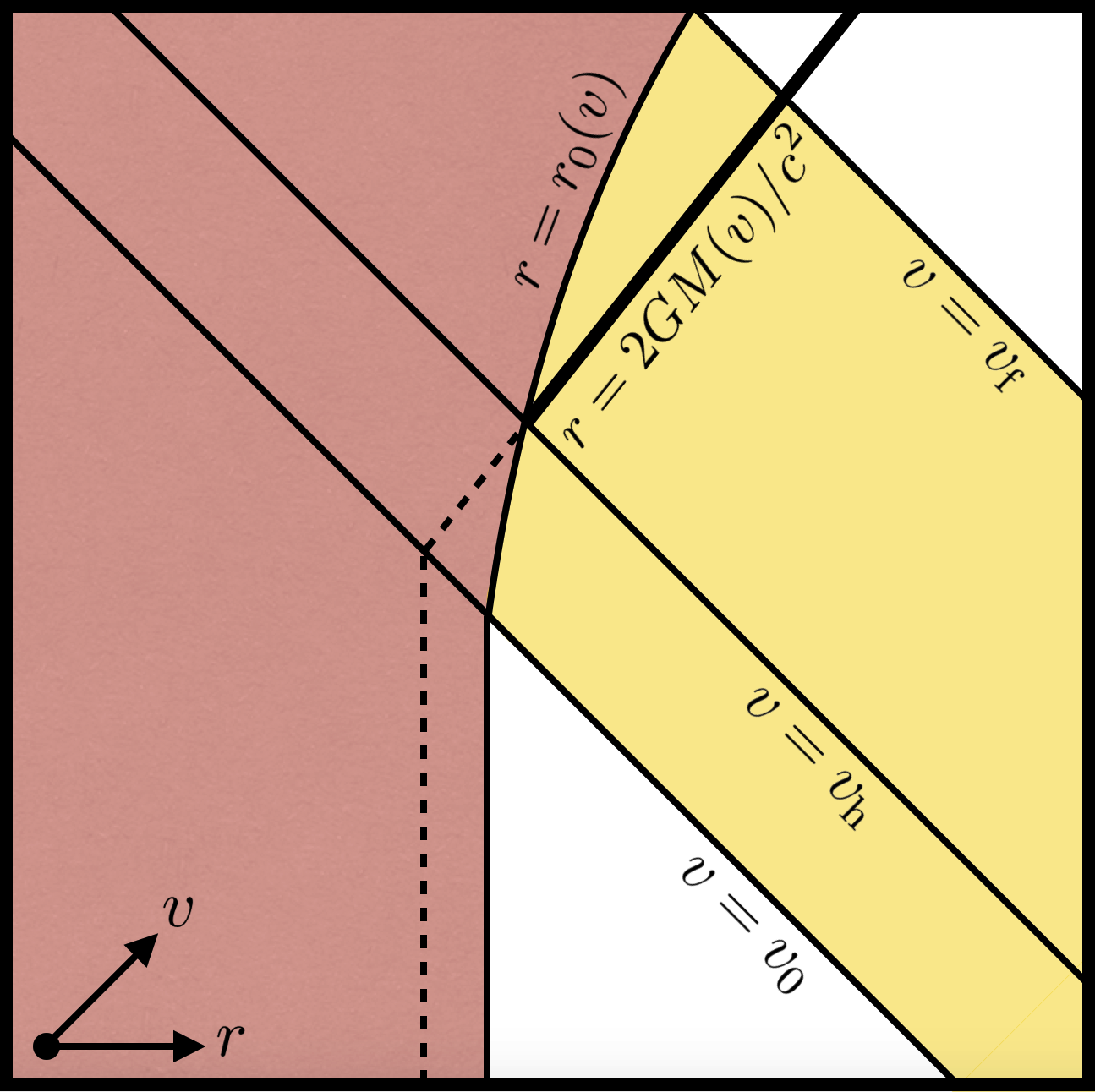}
\end{center}
\vspace{-0.5cm}
\caption{This figure shows the behavior under accretion of a HO with radius $r_0(v)$, and the subsequent formation of a trapping horizon. In this example, the surface $r_0(v)$ follows a time-like, but almost null, trajectory for $v<v_0$ and a null trajectory for $v\in[v_0,v_{\rm f}]$. The value of the function $2GM(v)/c^2$ is plotted for comparison, and it is given by the dashed line for $v<v_{\rm h}$ and the thick line for $v\geq v_{\rm h}$ (where it marks the position of the trapping horizon). The regions defined by $r> r_0(v)$ and $v< v_0$ or $v>v_{\rm f}$ (both in white) correspond to the Schwarzschild geometry, while the one defined by $r> r_0(v)$ and $v\in[v_0,v_{\rm f}]$ (in yellow) contains a patch of the Vaidya geometry. The region $r\leq r_0(v)$ (in red) describes the interior of the HO, the geometric details of which are not relevant for the present discussion.}
\label{fig:fig2}%
\end{figure}

The surface of the HO may also grow in time due to accretion. As discussed previously, the causality condition on this growth implies that the surface has to follow a time-like or, at most null, trajectory. For our purposes here it is enough to consider the limiting null case (see Fig. \ref{fig:fig2}), in which the growth proceeds at the maximum possible speed. Note that in this limiting case all the accreting matter is absorbed by the HO; when the trajectory of the HO is time-like, part of the accreting matter can be backscattered, following at most a null trajectory (see Fig. \ref{fig:fig1}). 

The maximum speed of growth is determined from Eq. \eqref{eq:vaidya} to be $dr_0/dv=[1-2GM(v)/c^2r_0(v)]/2$, which can be understood as a differential equation for $r_0(v)$. The initial value of this maximum speed is simply given by $dr_0/dv|_{v=v_0}=\mu/2$. This differential equation for $r_0(v)$ can be integrated for different values of the initial condition $dr_0/dv|_{v=v_0}$ (or equivalently, $\mu$). In particular, we have shown that if $dr_0/dv|_{v=v_0}<2G\dot{M}/c^3$, then $dr_0/dv<2G\dot{M}/c^3$ for all $v\geq v_0$, implying that the function $r_0(v)$ grows always slower than the function $2GM(v)/c^2$. As a result, the inequality \mbox{$r_0(v)\leq2GM(v)/c^2$} will be always satisfied for $v\geq v_{\rm h}$, with $v_{\rm h}$ some real number verifying
\begin{equation}
v_{\rm h}-v_0\leq 2\ell(4G\dot{M}/c^3-\mu)^{-1}.\label{eq:time}
\end{equation}
Hence, the external metric in Eq. \eqref{eq:vaidya}, which is glued with the internal geometry of the HO at its radius \mbox{$r=r_0(v)$}, displays a trapping horizon for $v\geq v_{\rm h}$ as long as $v_{\rm f}>v_{\rm h}$. For fixed $\ell$, the greater the accretion rate $\dot{M}$ the lower the value of the right-hand side of Eq. \eqref{eq:time} above. Thus for greater accretion rates or smaller values of $\ell$, local variations in time and space of the accretion rate become less important, which for the applications considered below justifies a posteriori the approximation of taking a constant and homogeneous accretion rate $\dot{M}$.

\section{Lower bounds}

For long enough accretion time \mbox{$(v_{\rm f}-v_0)/c$}, the formation of a trapping horizon can only be avoided if $dr_0/dv|_{v=v_0}\geq 2G\dot{M}/c^3$, which translates into the lower bound
\begin{equation}
\mu\geq \frac{4G\dot{M}}{c^3}.\label{eq:lbound}
\end{equation}
As stressed previously, this equation is strictly valid in a simplified scenario in which the HO is not rotating. However, we have estimated the effects of rotation and shown that this does not change significantly this lower bound: the right-hand side of equation (3) picks up a multiplicative factor that depends on the spin of the HO, but that is always of order 1 or greater. There is no reason to expect that relaxing the other simplifications regarding the nature of accreting matter will lead to important changes with respect to Eq. \eqref{eq:lbound}, as this must be in any case a relation between $\mu$ and the local flux of matter and energy across $r_0(v)$, $\dot{M}[r_0(v)]$. We stress that, as remarked in Table \ref{tab:1}, even changes of several orders of magnitude would not affect our main observational conclusions.

This lower bound is remarkably powerful. For Sgr A*, the accretion rate has a lower limit \mbox{$\dot{M}\gtrsim 10^{-11}\ M_\odot\ \mbox{yr}^{-1}$} given by the observed luminosity, so we obtain \mbox{$\mu\gtrsim 6\times10^{-24}$}. Therefore, all the theoretical models with $\mu\sim(L_{\rm P}/R_{\rm S})^\alpha$ and $\alpha\in[1, 2]$, which are not constrained by the upper bounds discussed previously (Table \ref{tab:1}), do not satisfy Eq. \eqref{eq:lbound}. The maximum time intervals in which a trapping horizon forms in these situations, as given by Eq. \eqref{eq:time}, are roughly $10^{-20}\ \mbox{s}$ and $10^{-64}\ \mbox{s}$ for $\alpha=1$ and $\alpha=2$, respectively, which are extremely short with respect to the typical timescales involved (e.g., the accretion timescale $t_{\rm acc}\simeq 4 \times10^8\ \mbox{yr}$). This justifies in particular the simplifying assumption that the accretion rate remains practically constant and homogeneous during these time intervals.

Higher accretion rates would lead to stronger lower bounds. For instance, active galactic nuclei (AGN) are powered by accretion and their luminosities reach (and sometimes significantly exceed) the Eddington luminosity \cite{Silverman2007,Yuan2014}, meaning the accretion rate is \mbox{$\dot{M}\geq L_{\rm Edd}/c^2\simeq 2 \times10^{26} M_9\ \mbox{g}\ \mbox{s}^{-1}$} (or $3 M_\odot\times M_9\ \mbox{yr}^{-1}$), where $M_9 = M/(10^9 M_\odot)$ is the mass of the central object. From Eq. \eqref{eq:lbound}, this accretion rate implies a lower bound $\mu\gtrsim 2 \times 10^{-12}\ M_9$. Much higher accretion rates are possible in systems where a BH merges with a neutron star (NS) or the core-collapse of very massive stars. The former case is believed to power short gamma-ray bursts (GRBs) and the latter is likely responsible for long GRBs \cite{Kumar2014}. Typical accretion rates for BH-NS mergers and long-GRB type core-collapses are $\dot{M}\sim1\ M_\odot\ \mbox{s}^{-1}$ and $\sim 0.1\ M_\odot\ \mbox{s}^{-1}$, respectively\footnote{We note that an alternative possibility for GRB central engine is a rapidly rotating (period $\sim 1\ \mbox{ms}$) and highly magnetized (surface magnetic field $\sim 10^{15}\ \mbox{G}$) neutron star, but BH-NS mergers must happen.}, which translates into $\mu \gtrsim 2 \times 10^{-5} [\dot{M}/(1\ M_\odot\ \mbox{s}^{-1})]$.

\begin{center}
\begin{table}[H]
\begin{center}
\begin{tabular}{|c|c|c|c|}
\hline
$\alpha$ &Theoretical $\mu$ &Lower bound on $\mu$ &Upper bound on $\mu$\\
\hline   
1 & $10^{-44}$ & \multirow{2}{*}{$10^{-24}$} & \multirow{2}{*}{$10^{-15}$}\\
\cline{1-2}   
2 & $10^{-88}$ & & \\
\hline   
\end{tabular}
\caption{Values of $\mu$ (only orders of magnitude) for Sgr A* and for two natural theoretical scenarios of HOs defined by different values of $\alpha$. These two cases were unconstrained by the previously known upper bound for Sgr A*, but both of them fail to satisfy the lower bound derived here. This lower bound (also the upper bound) is obtained under simplifying assumptions such as spherical symmetry or pressureless accreting matter. However, it is highly unlikely that including these additional features would change the lower bound by a large number of orders of magnitude (e.g., more than 64 for $\alpha=2$) so that the conclusions that are evident from the information in this table are modified.}
\label{tab:1}
\end{center}
\end{table}
\end{center}

\section{Conclusions}

We have found a novel mechanism, derived from the basic principle that interactions propagate at most with the speed of light, that permits to probe a previously unexplored region of parameter space for horizonless configurations. The main qualitative novelty with respect to previously known observational bounds is that this mechanism provides lower bounds on the radius of the horizonless object, reducing the parameter space still allowed by previous (upper) limits by up to $\sim 60$ orders of magnitude. We have provided estimations of the order of magnitude of these lower bounds for different astrophysical sources. Of particular importance are the robust consequences that follow for Sgr A*, as summarized in Table \ref{tab:1}. This makes stronger the observational case for the existence of regions in spacetime that behave locally as trapping horizons, which represents an important step forward in our understanding of the dynamical properties of astrophysical BHs.

\acknowledgments

The authors are grateful to Emil Mottola for many constructive exchanges on the subject of this work. R.C.-R. is grateful to Carlos Barceló, Luis J. Garay and Stefano Liberati for useful conversations, and also to the members of The Department of Astronomy at the University of Texas at Austin for their hospitality throughout the visit during which this work started to take shape. W.L. is funded by the Named Continuing Fellowship at the University of Texas at Austin.

\bibliography{refs}	

\end{document}